\documentclass[twocolumn,         % Format : preprint, twocolumn
               showpacs,            % Pacs : showpacs, noshowpacs
               preprintnumbers,     % Preprint: preprintnumbers,
               aps,                 % Society: ...
               prd,          	    % Journal Style : pra, prb, prc, prd, pre,
               a4paper,             % Size : a4paper, ...
               superscriptaddress,      % Affiliation (Title) : groupedaddress,
               nofootinbib,         % Footnote: footinbib, nofootinbib
               tightenlines,        % Remove additional spaces in a line
               floats,floatfix
               ]{revtex4-1}              
\usepackage{graphicx}  % needed for figures
\usepackage{dcolumn}   % needed for some tables
\usepackage{bm}   
\usepackage{amsmath,amssymb}
\usepackage{soul}
\usepackage{subfigure}
\usepackage{float}
\usepackage{makecell}
\usepackage[thinlines]{easytable}
\usepackage{array,booktabs}
\usepackage{lipsum} 
\usepackage{longtable}
\usepackage[colorlinks=true,linkcolor=blue,citecolor=blue]{hyperref}
\newcommand{\setone}{\textit{Pantheon Plus \ensuremath{+} CMB \ensuremath{+} DESI}}
\newcommand{\settwo}{\textit{DES Y5 + CMB + DESI}}
\newcommand{\setthree}{\textit{ CMB + DESI}}

\def\be{\begin{equation}}
\def\ee{\end{equation}}
\def\bea{\begin{eqnarray}}
\def\eea{\end{eqnarray}}

\def\d{{\rm d}}

\def\r{\right}
\def\l{\left}

\def\f{\frac}

\def\l{\left}
\def\r{\right}

\def\no{\nonumber}

\def\p{\phi}
\def\m{\rm m}
\def\pa{\partial}

\def\l{\left}
\def\r{\right}

\def\no{\nonumber}

\usepackage[colorinlistoftodos]{todonotes}
\newcommand{\addref}[1]{\todo[fancyline, color=red!50,size=\small]{Reference.}}
\begin{document} 

\title{Constraints on DBI dark energy with chameleon mechanism}

 \author{Burin Gumjudpai}\email{burin.gum@mahidol.ac.th}
  \affiliation{NAS, Centre for Theoretical Physics \& Natural Philosophy \\ Mahidol University, Nakhonsawan Campus,  Phayuha Khiri, Nakhon Sawan, Thailand 60130}
  \author{Nandan Roy} \email{nandan.roy@mahidol.ac.th}
  \affiliation{NAS, Centre for Theoretical Physics \& Natural Philosophy \\ Mahidol University, Nakhonsawan Campus,  Phayuha Khiri, Nakhon Sawan, Thailand 60130}
 \author{John Ward}\email{wardfunction@gmail.com}
  \affiliation{Arbutus Studios, West Broadway, Vancouver, BC, Canada, V6K0B1}

\date{\today}

\begin{abstract}

In this work, we investigate the Dirac--Born--Infeld (DBI) scalar field model with and without the inclusion of the chameleon mechanism in light of the latest cosmological observations. We constrain the model using data from Pantheon Plus, DES Y5, DESI DR2, and the compressed Planck likelihood. We consider an AdS throat of the form $f(\phi) = \lambda / \phi^4$ and a potential $V(\phi) = m_0^2 \phi^2 + m_1^2 \phi^4$. Our analysis shows that, for both cases, the mean value $m_1 \simeq 0$ suggests that the DBI field may lack significant self-interaction, with only an upper bound on $m_1$. The warp parameter is constrained to $\eta \geq 0$, while the chameleon coupling satisfies $\beta \leq 0$. No crossing of the phantom divide is observed under the assumed form of the warp factor and potential. We perform a statistical model comparison using the $\Delta \mathrm{AIC}$ relative to the $\Lambda$CDM model. Although the DBI model provides a slightly better fit to the data in terms of $\Delta \chi^2$, the improvement is negligible.  Consequently, the DBI model is mildly disfavored for both the cases compared to the $\Lambda$CDM model.

%We have found the fixed points and shown here that the late-time attractor solutions exist for these models. The work is done considering the general form of the potential and the warped factor.

\end{abstract}
\maketitle

 % \pacs{98.80.Cq}

\section{Introduction}
%%%%%%%%%%%%%%%%%%%%%%%

The simple solution to the problem of the present acceleration of the universe \cite{riess1998observational} is to postulate the existence of a vacuum energy or cosmological constant that agrees with all the current observational bounds \cite{Planck:2018vyg}. However, the vacuum energy needs to be fine-tuned to a small value on the other side to accommodate the present acceleration \cite{Padmanabhan:2002ji} and apart from this the vacuum energy or the cosmological constant must overcome the discrepancies coming from very recent cosmological observations \cite{Verde:2019ivm}. A discordance between the measurement of the Hubble parameter from the early
universe data, for example, Planck \cite{Planck:2018vyg}, BAO \cite{BOSS:2016wmc}, DES \cite{Joudaki:2019pmv} etc. and the late-time observations like SH0ES \cite{riess20162,riess2019large}, HOLiCOW \cite{wong2019h0licow}, CCHP \cite{freedman2019carnegie} etc. is of the order of $5 \sigma$.  Recent results derived from the Dark Energy Spectroscopic Instrument (DESI) \cite{DESI:2024mwx, DESI:2024aqx, DESI:2025zgx,DESI:2024jis,DESI:2025fii} and the Year 5 supernovae dataset from the Dark Energy Survey (DES) \cite{DES:2024jxu} indicate that at low redshift, both the BAO and supernovae data might not favor the conventional $\Lambda$CDM model, suggesting that dark energy could be dynamical in nature. These discrepancies between the observations and the theoretical problems faced by the cosmological constant, and the discordance in the value of the Hubble parameter, lead us to explore different alternative ideas for the cause of the accelerated expansion of the universe \cite{Copeland:2006wr, padmanabhan2006dark, amendola2010dark, clifton2012modified}.

Light scalar fields are frequently explored as possible contributors to the cosmic accelerating expansion \cite{Copeland:2006wr, clifton2012modified}. Various scalar field models, such as quintessence, k-essence and phantom field, have been developed to describe this phenomenon \cite{amendola2010dark, Bamba:2012cp}. In these scenarios, cosmic acceleration arises from the dynamics of a scalar field evolving under a potential, leading to an effective negative pressure that drives expansion \cite{Copeland:2006wr, clifton2012modified,  roy2022quintessence, Banerjee:2020xcn, Lee:2022cyh, Krishnan:2020vaf,Das:2019ixt}.

% Such fields naturally arise in UV-complete theories of quantum gravity and are expected to influence infrared (IR) phenomena like late-time acceleration. However, current candidates for quantum gravity do not yet provide a definitive explanation. Dynamical scalar fields can account for early-universe inflation and late-time dark energy. In particular, light moduli from string compactifications may couple to gravity and play this role \cite{conlon2006qcd}. In bottom-up approaches, scalar fields in low-energy effective field theories are introduced to explain acceleration without requiring a fully specified UV completion.

Although the introduction of scalar fields can help us to explain some outstanding problems in cosmology, they also create problems related to the local astronomical observations. Generally, the scalar field or the moduli considered in the cosmological models are massless. To be compatible with the local astronomical test, the masses of these scalar fields must be of the order of the Hubble scale or less.  These light scalar fields, if coupled to matter, would have fifth-force gravitational strength which could produce unacceptably large violation of the Equivalence Principle in the solar system or at smaller scales. Hence, a screening mechanism for the fifth force is needed in order for the theory to pass local experimental constraints, but allowing GR modification at large scales (see, e.g. \cite{Joyce2015} for review).

A screening mechanism called the chameleon mechanism can remove the mentioned problems of the light scalar field, that is, suppress the local fifth force effect \cite{Khoury2004}. In the chameleon mechanism, the mass of the scalar field depends on its ambient matter density. Specifically, the mass of the scalar field (the chameleon field) decreases in regions of low matter density, while it becomes larger in high-density regions. For a heavy chameleon field, the associated Compton wavelength is less, consequently the force mediated by the field becomes short-ranged \cite{Khoury2004a, Khoury2004b, Brax2004,Mota2007, Burrage2018}. In the early universe, the chameleon field faces significant challenges. As massive particle species become non-relativistic during the radiation era, the field experiences \textit{kicks} that induce rapid motion \cite{Brax:2004qh,Erickcek:2013dea,Erickcek:2013oma}. This can lead to a \textit{surfer} solution, where the field, moving at high velocity, tracks the minimum of its effective potential at constant Jordan-frame temperature. Upon reflection from the steep potential wall, the field's kinetic energy drops to zero, triggering sharp variations in its effective mass and exciting high-energy modes. As a result, classical evolution breaks down, undermining the chameleon model’s predictability during the Big Bang Nucleosynthesis epoch \cite{Banerjee2018, Damour1994}.

It has been shown that incorporating Dirac–Born–Infeld (DBI) corrections into chameleon models can mitigate the adverse effects of such \textit{kicks} in the early universe. The presence of the DBI term weakens the effective coupling between the chameleon field and matter preventing the field to reach very high velocity, thereby protecting the field from excitation of high-energy modes \cite{Padilla:2015wlv}. Hence introducing of the DBI field as the chameleon can avoid the surfer solution hence alleviating or solving the problem caused by chameleon at the Big Bang Nucleosynthesis.

The DBI models were first used in cosmology to explain inflation. The idea comes from string theory in which the open string in $Dp$-branes can be dynamical and drive inflation at the early epoch. The inflation field is dual to the radial coordinate of a $D3$ brane moving in a throat of warped compactified space. The motion of the $D3$ brane is limited in its speed, and it is affected by the speed of the brane and the throat's warp factor, which could be of AdS$_{5}$ type\cite{Silverstein:2003hf,Alishahiha:2004eh,Kecskemeti:2006cg,Chen:2004gc,Chen:2005fe,Ward:2007gs,Chen:2008hz,Gmeiner:2007uw,Gumjudpai:2009uy}. 
The kinetic part of the DBI action is a non-standard term and is a function of the scalar field, and the potential is identified with the compact manifold's local geometry, which is traversed to the $D3$-brane. In the context of inflation, the DBI field was initially suspected to generate excessively large non-Gaussianities. However, it was later shown that the simplest DBI inflationary models yield predictions similar to those of the standard scalar-field slow-roll scenarios \cite{Lidsey:2007gq, Baumann:2006cd, Alabidi:2008ej, Cheng:2013kpa, Huston:2008ku}. Subsequently, DBI models have been explored as potential candidates for dark energy \cite{Ahn:2009xd, Garousi:2004ph, Martin:2008xw, Ahn:2009hu, Copeland:2010jt, Rezazadeh:2020zrd, Gumjudpai:2009uy}. More recently, the DBI field has also been proposed as a possible candidate for dark matter, particularly in relation to addressing the $S_8$ tension \cite{Suo:2023xes}.

In this work, we consider the DBI scalar field as the dark energy component of the universe and employ current cosmological observations, including Supernovae, Baryon Acoustic Oscillations (BAO), and Cosmic Microwave Background (CMB) data, to constrain the model parameters. We examine two distinct scenarios: the DBI field without the chameleon mechanism and the DBI field incorporating the chameleon mechanism.

The manuscript is organized as follows. In Section II, we present the derivation of the field equations from the DBI action. Section III describes the observational data sets used in this work. The results of our analysis are discussed in Section IV, and we summarize our conclusions in Section V.

%\todo[inline,color=yellow!40]{Discussion about the dynamical systems analysis.} 

%%%%%%%%%%%%%%%%%%%%%%%
\section{DBI action with chameleon mechanism}
%%%%%%%%%%%%%%%%%%%%%%%

Considering the action for a BPS $D3$-brane localized
in a warped compacification of type IIB string theory. We will also assume the existence
of a matter sector coupled to the world-volume theory of the brane. The resulting action can then
be written in the form \cite{Gumjudpai:2009uy}
\begin{equation}
\begin{split}
S_{\phi} &= \\ 
- & \int \d^4 x \sqrt{-g} \l( T(\phi) \l( W(\phi) \sqrt{1-\f{2 X}{T}} - 1\r) + V(\phi)\r)   \end{split} \label{DBIaction}
\end{equation}
This is the generalised Dirac-Born-Infeld (DBI) action where
\bea
X  &\equiv&  -\f{1}{2} g^{\mu\nu} \pa_{\mu}\phi \pa_{\nu} \phi   =  -\f{1}{2}(\nabla \phi)^2    \no\\
T& \equiv &1/f(\phi)   \no
\eea
where \( f(\phi) \) denotes the warp factor, and since we take \( W = 1 \), this corresponds to the standard DBI case. One can find that the action in Eq.(\ref{DBIaction}) reads
\begin{equation}
S_{\phi} = \int \d^4 x \sqrt{-g} \Big[ -T \l(\Gamma -1  \r)  - V \Big]
\end{equation}
where $\Gamma  \equiv  \sqrt{1+\f{(\nabla \phi)^2}{T}}$. In this analysis, we take the DBI warp factor to have an anti–de Sitter (AdS) profile,

\begin{equation}
    f(\phi) = \frac{\eta}{\phi^4},
\end{equation}

where $\eta$ is a dimensionless constant parameter that determines the brane tension within the DBI setup.

We define the DBI Lagrangian density as $\mathcal{L}_{\p} \equiv  -T \l(\Gamma -1  \r)  - V$.
In this system the total action is of the Einstein-Hilbert, matter field and the DBI scalar field terms,
\begin{equation}
\begin{split}
S = \int \mathrm{d}^4 x \bigg\{ \mathcal{L}_{\mathrm{EH}} &+ \mathcal{L}_{\mathrm{m}}(\psi_{\mu}, \tilde{g}_{\mu\nu}) + \mathcal{L_{\phi}} \bigg\},
\end{split}
\label{eq_action}
\end{equation}

% This can read
%\begin{equation} 

%S = \int \d^4 x \sqrt{-g} \left( \frac{M_{\rm Pl}^2}{2}\mathcal{R} - T(\phi) (\Gamma -1) - V(\phi) +  \frac{1}{\sqrt{-g}} \mathcal{L}_{\m}(\psi_{\m}, \tilde g_{\mu \nu})\right)
%\end{equation}
where ${\mathcal{L}}_{\rm EH}  \equiv ({M_{\rm Pl}^2}/{2})\sqrt{-g} \mathcal{R}$ and
${\mathcal{L_{\phi}}} = \sqrt{-g} \l[ -T \l(\Gamma -1  \r)  - V \r] $.
The $\mathcal{R}$ is Ricci-scalar, $M_{\rm Pl}$ is reduced Planck mass, and $\psi_{\mu}$ is a matter field. We define Jordan frame metric
\be
 \tilde g_{\mu \nu} = e^{2 \beta \phi/M_{\rm Pl}} g_{\mu \nu}.   \label{chameleon mech}
 \ee
  The kinetic term
for the scalar field $\phi$ is encoded in the DBI-part of the action ($-T \l(\Gamma -1  \r)$).  The coupling constant $\beta$ can take either positive or negative values. A negative coupling simply indicates that the Jordan-frame metric scales with $\phi$ in the opposite direction compared to the positive case. The sign choice does not introduce any inconsistency into the chameleon framework, since the existence of a density-dependent minimum in the effective potential only requires that $V_{,\phi}$ and $\beta$ have opposite signs \cite{Khoury:2003rn}.

%Variation of the action with respect to the scalar field to get  the Euler-Lagrange equation
%\be
%\f{\delta  \mathcal{L}_{\phi}}{\delta \phi } \, =\,  \f{\pa \mathcal{L}_\phi }{\pa \phi }  -   \nabla_{\mu} \l[ \f{ \pa \mathcal{L}_\phi  }{\pa (\nabla_{\mu} \phi)  }   \r]   =  0
%\ee
Matter field term can be varied independently to have $\delta S_{\m} = 0$, into the total action. Symbol $'$ denotes $\d/\d \phi$. Following details in the appendix,
we  obtain the DBI equation of motion,
\be
\begin{split}
 \f{\Box^2 \phi }{\Gamma}  \,  -\f{T'}{2\Gamma} \, (\Gamma -1)^2  \,  -  \, \f{1}{\Gamma^2} \, g^{\mu\nu} (\pa_{\mu} \phi )(\pa_{\nu} \Gamma) \\
 \:=\:   \, V'  - \f{\mathcal{L}'_m}{\sqrt{-g}}
 \end{split}
\ee
The last term is the matter Lagrangian density. With coupling between the scalar field to the metric via the equation (\ref{chameleon mech})-the chameleon mechanism, the matter Lagrangian reads
\be
\mathcal{L}'_{\m}  \;=\; - \sqrt{-g} \f{\beta}{M_{\rm Pl}} \,\rho(1-3w)\, e^{\beta(1-3w) \phi /M_{\rm Pl}}
\ee
where $w$ is equation of state parameter of barotropic fluid, i.e. matter and radiation. Considering dust fluid ($w=0$), the full DBI field equation of motion with chameleon mechanism is hence
\be \label{eq:eom}
\begin{split}
 \f{\Box^2 \phi }{\Gamma}  \,  -\f{T'}{2\Gamma} \, (\Gamma -1)^2  \,  -  \, \f{1}{\Gamma^2} \, g^{\mu\nu} (\pa_{\mu} \phi )(\pa_{\nu} \Gamma)  \;\\
 \:=\:\;   \, V'              +  \f{\beta}{M_{\rm Pl}} \,\rho \, e^{\beta \phi /M_{\rm Pl}}  
 \end{split}
\ee
The last term on the right hand side is proportional to the energy density of the matter sector. One finds this expression by noting that variation of
the matter Lagrangian yields a term proportional to $\tilde T^{\mu \nu} \tilde g_{\mu \nu}$ which for an isotropic fluid will be of the form $-(1-3w)\tilde \rho$ in
the Jordan frame. In Einstein frame we see that $\rho = \tilde \rho\, e^{3(1+ w) \beta \phi/M_{\rm Pl}}$.
The effective potential is therefore defined as
\begin{equation}
V'_{\rm{eff}}  \;\;\equiv\; \;V'+ \frac{\beta \rho}{M_{\rm Pl}} e^{ \beta \phi/M_{\rm Pl}}.
\end{equation}
%The notation we use is that $\partial_{X} = X_{\phi}$ for
%a quantity $X$, and $\Box^2$ has the usual form in a curved geometry.
%%
The field equation  (\ref{eq:eom}) can be treated as temporal dependent and radial dependent. Their solutions will be investigated here.

%%%%%%%%%%%%%%%%%%%%%%%

Assuming that the metric is flat FRW and that the scalar field is homogeneous, i.e. only-time dependent, the equation of motion therefore reduces to
\be
\begin{split}
- \ddot{\phi} - \Gamma^2\, 3 H \dot{\phi}  + \f{T'}{2} \l(1 - 3 \Gamma^2  + 2\Gamma^3 \r)  \;\\
\,=\;\, \l( V'    +  \f{\beta}{M_{\rm Pl}} \,\rho \, e^{\beta \phi /M_{\rm Pl}} \r) \Gamma^3    \label{tempfield}
\end{split}
\ee
where here $\Gamma = (1 - \dot{\phi}^2/T)^{1/2}$.

The density and the pressure of the DBI field can be written as;
\begin{subequations}
\begin{eqnarray}
p_{\phi} &=& T(\phi)\Gamma(\phi) [1/\Gamma(\phi) - 1] - V(\phi) \, ,\\
\rho_{\phi} &=& T(\phi)[1/\Gamma(\phi) -1] + V(\phi) \, .
\end{eqnarray} \label{eq:pressure}
\end{subequations}

To determine the potential governing the dynamics of the DBI scalar field in our dark energy scenario, the action (\ref{DBIaction}) must remain invariant under a group of internal symmetries. In particular, for the action to be symmetric under the transformation $\phi \to -\phi$, the potential $V(\phi)$ must be an even function. Following previous studies \cite{Ahn:2009xd, Rezazadeh:2020zrd}, we adopt the form

\begin{equation}
  V(\phi) = m_0^2 \phi^2 + m_1^2 \phi^4,  
\end{equation}

where $m_0$ and $m_1$ are constant parameters. The parameter $m_0$ sets the effective mass scale of the scalar field and has mass dimension one, whereas $m_1$ is dimensionless and governs the strength of the quartic self-interaction.

 This potential structure naturally accommodates spontaneous symmetry breaking at an energy scale significantly higher than that of the present universe—a feature with important implications for theoretical cosmology \cite{Liddle:2000cg}.

\section{Observational DATA}

In this section, we investigate the observational constraints on the model by comparing its predictions with current cosmological data, considering both cases: with and without the inclusion of the chameleon mechanism. We modify the publicly available code \texttt{CosmoDS}~\cite{Roy:2026icy} to implement the background equations and employ \texttt{Cobaya}~\cite{Torrado:2020dgo} to perform the Markov Chain Monte Carlo (MCMC) analysis. The resulting posterior distributions are visualized using the \texttt{GetDist} plotting package. The \texttt{CosmoDS} code is used as an external cosmology module for \texttt{Cobaya}, allowing the likelihoods already implemented in \texttt{Cobaya} to be readily utilized. The convergence criterion we adopt here is $R - 1 \approx 0.02$. A detailed description of the datasets used in this study is provided below.

\subsection{Supernova Data}
Type Ia supernovae are commonly utilized as standard candles due to their relatively uniform intrinsic luminosity~\cite{reiss1998supernova, SupernovaSearchTeam:1998fmf}. In this work, we have used the Pantheon Plus compilation of SN Ia data~\cite{Scolnic:2021amr, Riess_2022, Brout:2022vxf, Riess:2021jrx} and the DES Year 5 dataset~\cite{DES:2024jxu}. These datasets are based on distinct photometric systems and selection criteria, and both provide measurements of the distance modulus $\mu$ across a wide range of redshifts.

\subsection{DESI BAO Data}

The density distribution of visible baryonic matter exhibits periodic fluctuations, known as baryon acoustic oscillations (BAO), which serve as important standard rulers for precise distance measurements in cosmology. In this work, we make use of the 2025 BAO observational data from the Dark Energy Spectroscopic Instrument (DESI-DR2), as presented in~\cite{DESI:2025zgx, DESI:2024mwx}. 

The BAO measurements provide constraints on effective distance scales both along the line of sight and in the transverse direction. These are defined through the following relations.

For the line-of-sight direction, the comoving Hubble distance is given by:
\begin{equation}
\frac{D_H(z)}{r_d} =\frac{c r_d^{-1}}{H(z)},
\end{equation}
where $c$ is the speed of light, $H(z)$ is the Hubble parameter at redshift $z$, and $r_d$ is the comoving sound horizon at the drag epoch.

For the transverse direction, the comoving angular diameter distance is expressed as:
\begin{equation}
    \frac{D_{M}(z)}{r_{d}}\equiv\frac{c}{r_{d}}\int_{0}^{z}\frac{d\tilde{z}}{H(\tilde{z})}=\frac{c}{H_{0}r_{d}}\int_{0}^{z}\frac{d\tilde{z}}{h(\tilde{z})}.
\end{equation}
where $H_0$ is the present-day Hubble parameter and $h(z) = H(z)/H_0$.

The spherically averaged effective distance is defined as:
\begin{equation}
\frac{D_V(z)}{r_d} =\left[\frac{c z r_d^{-3} d_L^2(z)}{H(z)(1+z)^2}\right]^{\frac{1}{3}}.
\end{equation}
where $d_L(z)$ is the luminosity distance at redshift $z$.

In this study, we determine the comoving sound horizon $r_d$ via the fitting formula \cite{Brieden:2022heh,DESI:2025zgx, DESI:2024mwx}, assuming standard pre-recombination physics,

\begin{equation}
r_d = 147.05\,\mathrm{Mpc} \times 
\left( \frac{\omega_b}{0.02236} \right)^{-0.13}
\left( \frac{\omega_{bc}}{0.1432} \right)^{-0.23}
\left( \frac{N_{\mathrm{eff}}}{3.04} \right)^{-0.1},
\end{equation}

where $\omega_b \equiv \Omega_b h^2$ and $\omega_{bc} \equiv (\Omega_b + \Omega_c) h^2$.
\subsection{Compressed CMB Likelihood}

We also utilize the Planck 2018 compressed likelihood, following the methodology presented in~\cite{Chen:2018dbv} (hereafter referred to as CMB). The compressed CMB prior provides a practical alternative to performing a full likelihood analysis of the complete Planck dataset, especially when exploring dark energy models that extend beyond the standard $\Lambda$CDM framework. This approach retains a comparable constraining power to the full Planck data while significantly reducing computational complexity.

The compressed likelihood incorporates the physical baryon density $\omega_b = \Omega_b h^2$, along with two key shift parameters: $l_{\mathrm{A}} = (1+z_*) \frac{\pi D_{\mathrm{A}}(z_*)}{r_s(z_*)}$ and $\mathcal{R} = \sqrt{\Omega_M H_0^2}\, D_A(z_*)$, where $z_*$ denotes the redshift at decoupling and $D_A$ is the comoving angular diameter distance. It is important to note, however, that these priors may not fully capture the richness of the complete CMB power spectrum, particularly in models that exhibit significant deviations from $\Lambda$CDM. In such cases, systematic biases in the inferred cosmological parameters may arise~\cite{Corasaniti:2007rf, Elgaroy:2007bv}.

In this work, we consider the following three combinations of datasets to constrain our model:

\begin{enumerate}
\item \setone
\item \settwo
\item \setthree
\end{enumerate}

\section{Results}

\subsection{Without chameleon mechanism}

Here, we study the DBI model without the chameleon mechanism, hence $\beta = 0$. We have considered a flat prior on the cosmological parameters  $\Omega_{m0}:[0,0.4], H_0:[60,80], \Omega_b:[0.01,0.1]$. The prior on the parameter related to the warp factor  $\eta:[-5,5]$, and on the potential parameters, $m_0:[0,20], m_1:[-20,20]$. Since $m_0$ denotes the effective mass of the scalar field, we impose the prior condition $m_0>0$.
We adopt a prior on the present-day value of the scalar field and its velocity, $\phi$ and $\dot{\phi}$, such that $\phi_0 \in [0.5,1.5]$ and $\dot{\phi}_0 \in [0.005,0.015]$. In the DBI setup, the Lorentz factor is defined as $\gamma = \sqrt{1 - f(\phi)\dot{\phi}^2}$, which under our parametrization becomes $\gamma = \sqrt{1 - \dot{\phi}^2/T}$ with $T = \phi^4/\eta$. For the theory to be mathematically consistent, the expression inside the square root must be positive, leading to the requirement $\dot{\phi}^2 < \phi^4/\eta$. This inequality determines the physically allowed region of the DBI scalar field phase space. Our choice of prior on $\phi_0$ and $\dot{\phi}_0$ is to guarantee that the initial conditions lie within this admissible region and that the DBI Lorentz factor remains real throughout the entire cosmological evolution.

\begin{table*}
\centering
\resizebox{\textwidth}{!}{
\begin{tabular}{|l|lll|lll|lll|}
\hline
\toprule
Parameters & \multicolumn{3}{c |}{\setone} & \multicolumn{3}{c |}{\settwo} & \multicolumn{3}{c |}{\setthree} \\
 \hline
 & $\Lambda$CDM & DBI & DBI (Chameleon) & $\Lambda$CDM & DBI & DBI (Chameleon) & $\Lambda$CDM & DBI & DBI (Chameleon) \\
\midrule
$H_0$ & 68.3086 ± 0.2990 & 68.2883 ± 0.2993 & 68.3030 ± 0.2927 & 68.2110 ± 0.2941 & 68.1888 ± 0.2907 & 68.1747 ± 0.2962 & 68.4339 ± 0.3049 & 68.4034 ± 0.3005 & 68.4079 ± 0.3080 \\
$\Omega_m$ & 0.3010 ± 0.0038 & 0.3012 ± 0.0038 & 0.3011 ± 0.0037 & 0.3022 ± 0.0038 & 0.3025 ± 0.0038 & 0.3027 ± 0.0038 & 0.2993 ± 0.0039 & 0.2997 ± 0.0039 & 0.2997 ± 0.0039 \\
$\Omega_b$ & 0.0479 ± 0.0003 & 0.0479 ± 0.0003 & 0.0479 ± 0.0003 & 0.0480 ± 0.0003 & 0.0480 ± 0.0003 & 0.0480 ± 0.0003 & 0.0478 ± 0.0003 & 0.0478 ± 0.0003 & 0.0478 ± 0.0003 \\
$\beta$ & — & — & $< -0.142                  $ & — & — & $< -0.146                  $ & — & — & $< -0.150                  $ \\
$\eta$ & — & $> 0.245$  & $> 0.288$ & — & $> 0.319                   $ & $> 0.310                   $ & — & $> 0.295 $ & $> 0.0558           $ \\
$m_0$ & — & $< 9.60$  & $< 31.5                    $ & — & $< 9.81                    $ & $< 31.1 $ & — & $< 9.43                    $ &  $< 31.0                    $ \\
$m_1$ & — & -0.0456 ± 1.6573 & 0.1172 ± 15.4833 & — & 0.0130 ± 1.6610 & -0.0425 ± 16.1462 & — & 0.0172 ± 1.6312 & 0.5601 ± 15.4006 \\
$\chi^2$ & 1419.89 & 1419.56 & 1419.56 & 1663.83 & 1663.31 & 1663.31 & 13.81 & 13.57 & 13.58 \\
$\Delta \chi^2$ & 0.00 & -0.33 & -0.33 & 0.00 & -0.52 & -0.52 & 0.00 & -0.24 & -0.23 \\
$\Delta \mathrm{AIC}$ & 0.00 & 2.67 & 3.67 & 0.00 & 2.48 & 3.48 & 0.00 & 2.76 & 3.77 \\
\bottomrule
\hline
\end{tabular}}
\caption{Mean values of different cosmological parameters together with $68\%$ constraints for the DBI model with and without the Chameleon mechanism for all three data sets combinations. The result for the $\Lambda$CDM model is reported for comparison.}\label{tab:constaint}
\end{table*}

\begin{figure}
    \centering
    \includegraphics[width=1\linewidth]{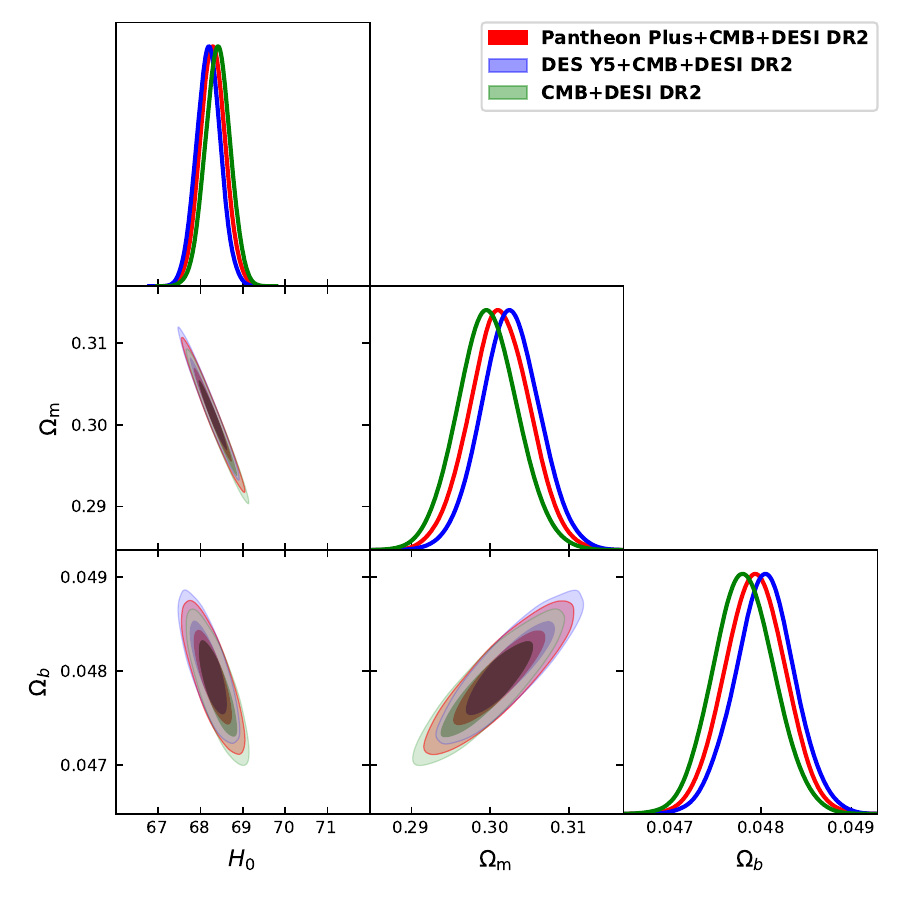}
    \caption{The plot displays the triangular representation of the 1D and 2D posterior distributions for various cosmological parameters in the DBI model without the chameleon mechanism. The red contours correspond to results from \setone, blue to \settwo, and green to \setthree. }
    \label{fig:cosmo_no}
\end{figure}

\begin{figure}
    \centering
    \includegraphics[width=1\linewidth]{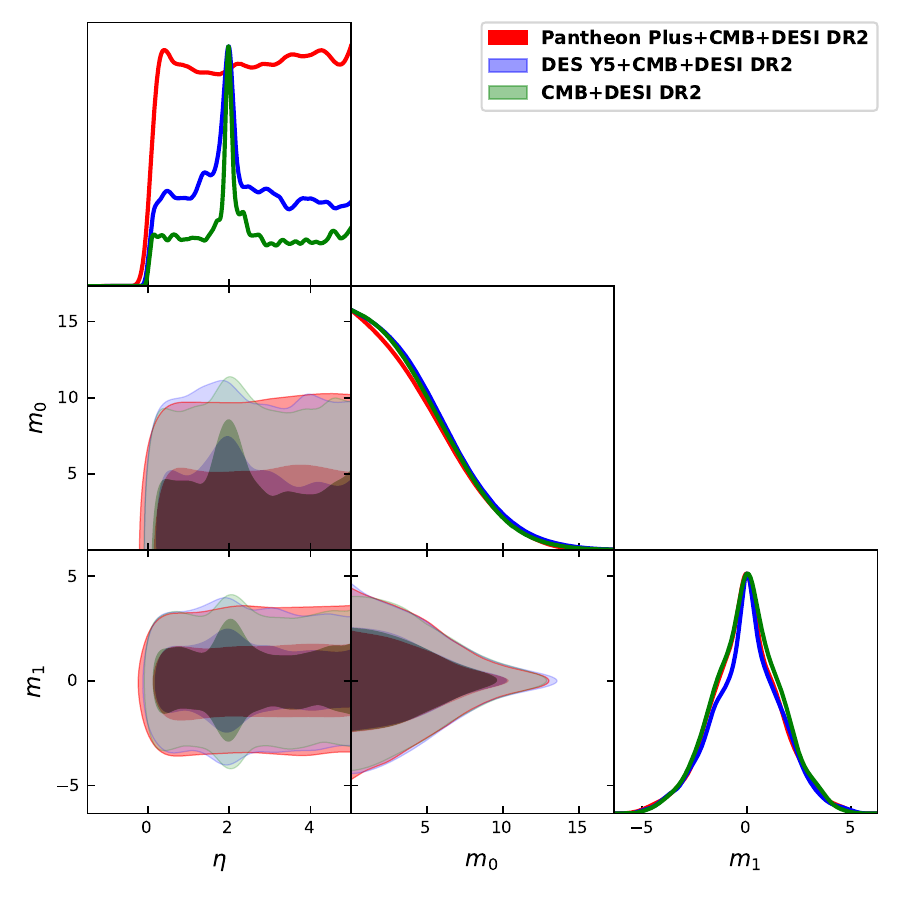}
    \caption{The plot displays the triangular representation of the 1D and 2D posterior distributions for model parameters $\eta, V_0, V_1$ in the DBI model without the chameleon mechanism. The red contours correspond to results from \setone, blue to \settwo, and green to \setthree.}
    \label{fig:model_no}
\end{figure}

The constraint derived from our analysis is presented in Table \ref{tab:constaint}, along with the mean and $1\sigma$ constraint level on the cosmological parameters.  In Fig.\ref{fig:cosmo_no} we have shown the $1D$ and $2D$ triangular plot of the posterior distribution of $H_0, \Omega_m, \Omega_b$ obtained from our analysis. The red plot depicts the posteriors obtained for the \textit{\setone} data combination, the blue corresponds to the \textit{\settwo}, and for the \textit{\setthree} it is shown in green. Figure \ref{fig:cosmo_no} illustrates that the posteriors derived from each of these data sets exhibit remarkable consistency. The plot of the constraint obtained for the model parameters $\eta, m_0, m_1$ is shown in Fig.\ref{fig:model_no}. From our analysis, the scalar field throat parameter $\eta$ has a lower bound approximately $eta > 0.24$ up to $3\sigma$ for all three data sets. The parameter $m_0$ has an upper bound, whereas the obtained constraint on the mean value indicates $m_1 \simeq 0$. This suggests that the self-interaction of the DBI field could potentially be negligible.

In this work, we analysed the statistical performance of DBI model in comparison to the standard $\Lambda$CDM cosmology. This evaluation was conducted using the minimum chi-squared statistic ($\chi^2_{\text{min}}$), the chi-squared difference ($\Delta \chi^2$), and the difference in the Akaike Information Criterion ($\Delta \text{AIC}$). Specifically, we computed the difference in minimum chi-squared values, $\Delta \chi^2_{\text{min}}$, between our model and the $\Lambda$CDM baseline, with the results presented in Table~\ref{tab:constaint}. For each dataset considered, the DBI model without the chameleon mechanism slightly better fits the data than the $\Lambda$CDM framework.

As shown in Table~\ref{tab:constaint}, for \textit{\settwo}, the minimum chi-squared value is the lowest for the DBI model, yielding a difference of $\Delta \chi^2_{\text{min}} = -0.52$ relative to $\Lambda$CDM. For \textit{\setone}, the difference is $\Delta \chi^2_{\text{min}} = -0.33$, and for \textit{\setthree}, it is $\Delta \chi^2_{\text{min}} = -0.23$. These results indicate that the DBI model can fit the data very similarly to the $\Lambda$CDM model.

However, to determine the preference of a model, one needs to take into account the number of free parameters a theory has. To find it out, we have computed the difference in AIC values using the following relation:
\begin{equation}
\Delta \text{AIC} = \chi^{2}_{\text{min}, \mathcal{M}} - \chi^{2}_{\text{min}, \Lambda \text{CDM}} + 2 (N_{\mathcal{M}} - N_{\Lambda \text{CDM}}),
\end{equation}
where $\mathcal{M}$ represents the model being tested, and $N_{\mathcal{M}}$ and $N_{\Lambda \text{CDM}}$ denote the number of parameters in models $\mathcal{M}$ and $\Lambda$CDM, respectively. A negative $\Delta \text{AIC}$ relative to $\Lambda$CDM indicates that the model $\mathcal{M}$ is statistically preferred. The AIC criterion penalizes models with greater complexity, accounting for the number of free parameters. According to standard AIC interpretation, differences between $-2$ and $-4$ suggest weak evidence in favor of the model, differences between $-4$ and $-7$ indicate moderate support, and differences less than $-10$ imply strong support over $\Lambda$CDM.

Applying this framework to the datasets \textit{\setone}, \textit{\settwo}, and \textit{\setthree}, we find $\Delta \text{AIC}$ values of $2.67$, $2.48$, and $2.76$, respectively. These findings indicate that the $\Lambda$CDM model remains the preferred model across all three combinations of datasets.

\begin{figure}
    \centering
    \includegraphics[width=1\linewidth]{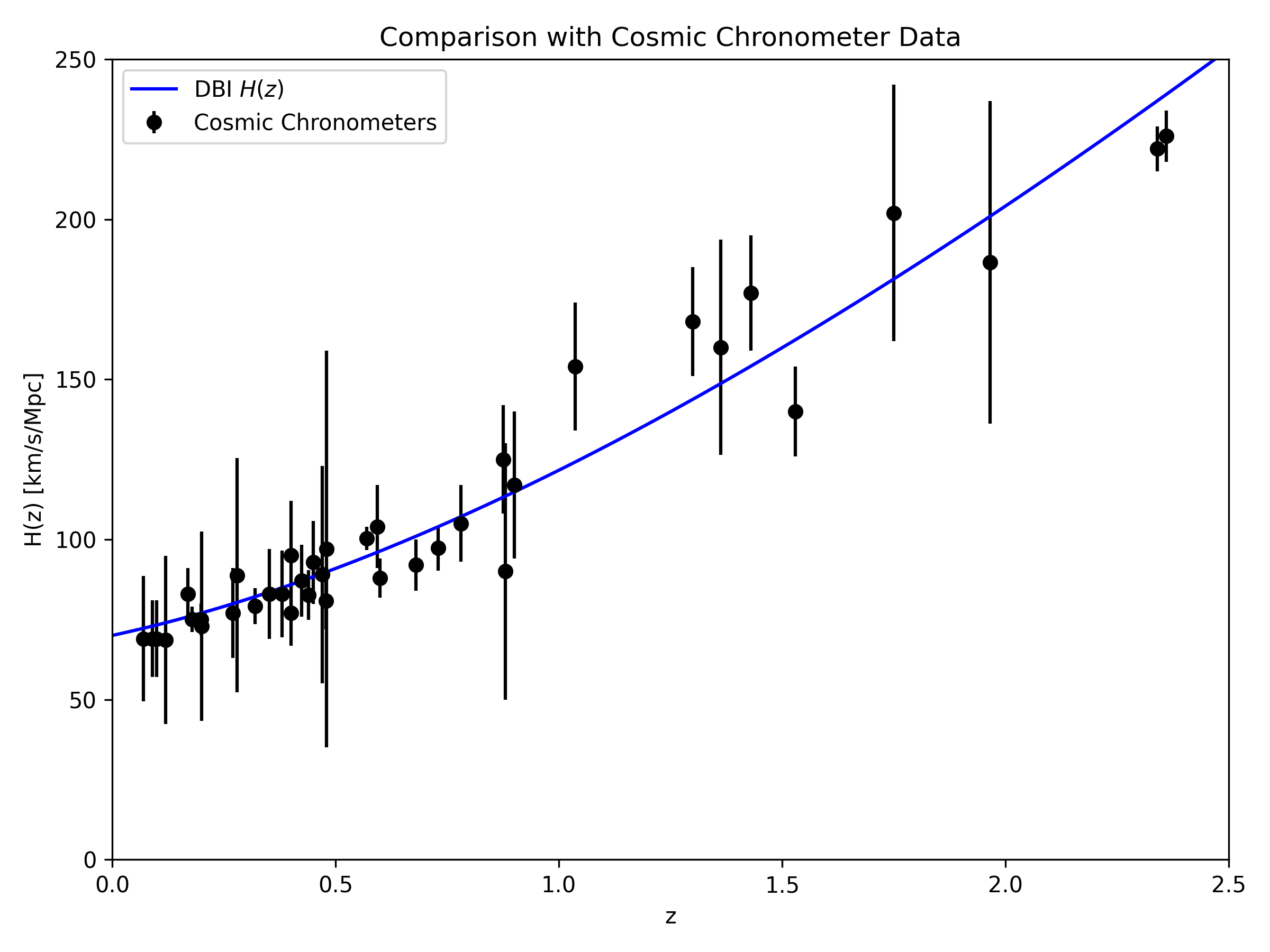}
    \caption{Evolution of 
$H(z)$ versus redshift
$z$ for the DBI model without the chameleon mechanism. Data points from Cosmic Chronometer datasets are shown for comparison. For the choice of the parameters $m_0 =0.1, m_1 = 10^{-4},\eta=1$ with the current value of the $\phi_0 \simeq 1, \dot{\phi}_0 \simeq 0.001$.}
    \label{fig:Hz_no}
\end{figure}

\begin{figure}
    \centering
    \includegraphics[width=1\linewidth]{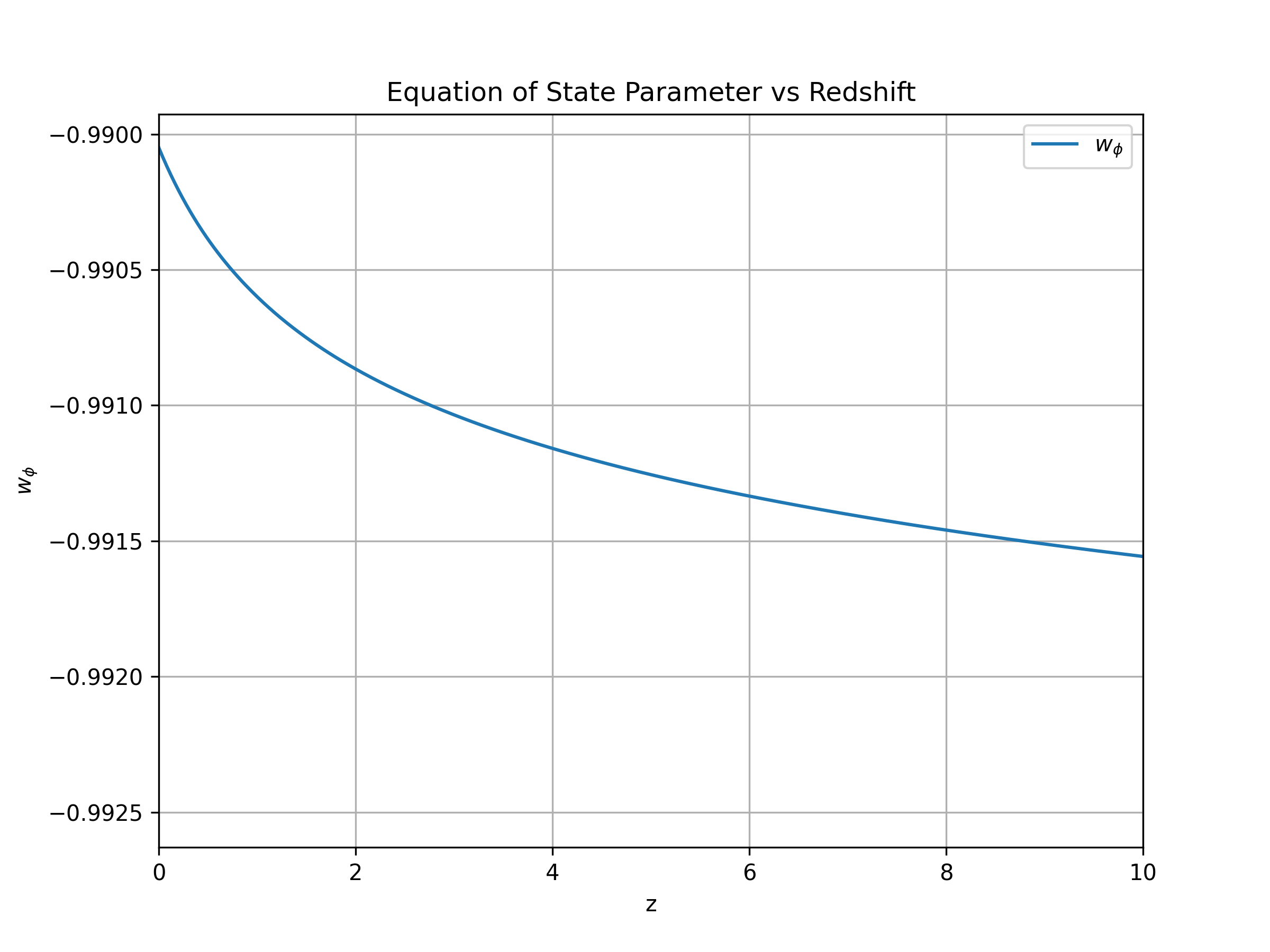}
    \caption{Evolution of the EOS of the dark energy for the DBI model without the chameleon mechanism. For the choice of the parameters $m_0 =0.1, m_1 = 10^{-4},\eta=1$ with the current value of the $\phi_0 \simeq 1, \dot{\phi}_0 \simeq 0.001$.}
    \label{fig:wphi_no}
\end{figure}

To test the evaluation of the cosmological parameters, we have plotted the evolution of the $H(z)$ against $z$ in Fig. \ref {fig:Hz_no}. The observational data from the cosmic chronometer data set has also been plotted for comparison. The choice of the parameter value for this plot is within the range of the posteriors obtained from the MCMC analysis. The plot of the evolution of the EoS of the scalar field for the same set of parameters is shown in the Fig.\ref{fig:wphi_no}. The EoS is very close to the $w_\phi = -1$ and shows a slight deviation from it at late times.

\subsection{With chameleon mechanism}

Here, we study the DBI model with the chameleon mechanism activated, hence $\beta \neq 0$. We have considered a flat prior on the cosmological parameters  $\Omega_{m0}:[0,0.4], H_0:[60,80], \Omega_b:[0.01,0.1]$. The prior on the parameter related to the warp factor is $\eta:[-5,5]$, the chameleon parameter $\beta : [-1.0, 1.0]$, the priors on the potential parameters are $m_0:[0,40], m_1:[-40,40]$ and $\phi_0 \in [0.5,1.5], \dot{\phi}_0 \in [0.005,0.015]$. The constraints obtained from our analysis are reported in Table \ref{tab:constaint}, along with the mean and 68\% constraints on the cosmological parameters.

\begin{figure}
    \centering
    \includegraphics[width=1\linewidth]{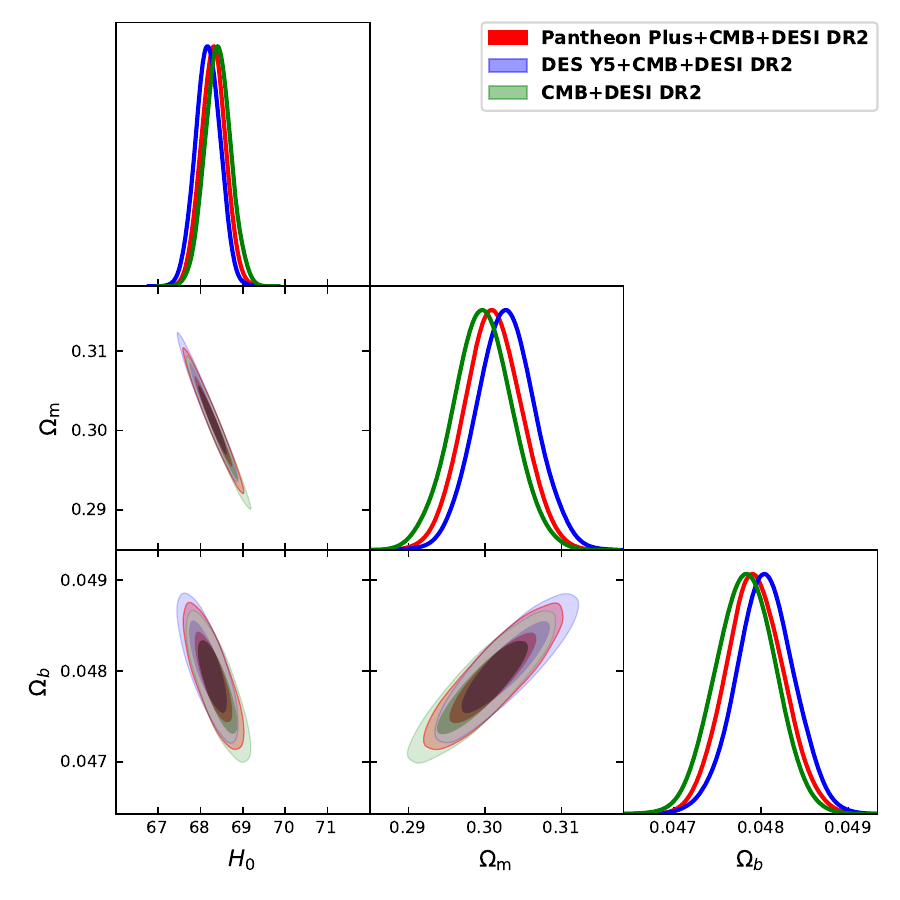}
    \caption{The plot displays the triangular representation of the 1D and 2D posterior distributions for various cosmological parameters in the DBI model with the chameleon mechanism. The red contours correspond to results from \setone, blue to \settwo, and green to \setthree.}
    \label{fig:cosmo_with}
\end{figure}

\begin{figure}
    \centering
    \includegraphics[width=1\linewidth]{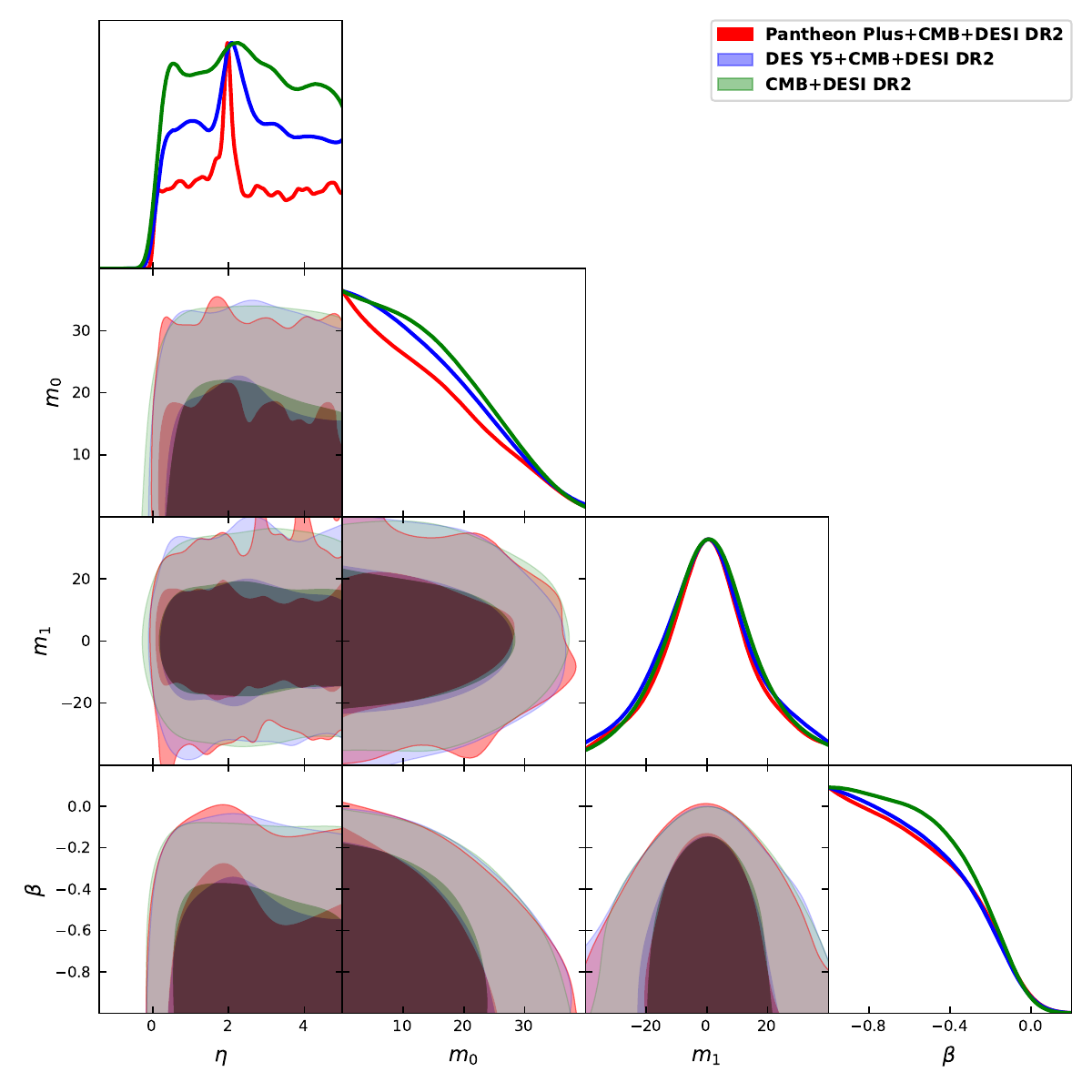}
    \caption{The plot displays the triangular representation of the 1D and 2D posterior distributions for model parameters $\eta, V_0, V_1$ in the DBI model with the chameleon mechanism. The red contours correspond to results from \setone, blue to \settwo, and green to \setthree.}
    \label{fig:model_with}
\end{figure}

Figure \ref{fig:cosmo_with} presents the one-dimensional and two-dimensional triangular plots of the posterior distributions for $H_0,  \Omega_m, \Omega_b$ as derived from our analysis. The posterior corresponding to the \textit{\setone} data set is illustrated in red, the \textit{\settwo} in blue, and the \textit{\setthree} in green. Meanwhile, Figure \ref{fig:model_with} displays the constraints for the model parameters $\eta, \beta, m_0, m_1$. Similar to the previous case, in this scenario of the DBI field with the chameleon mechanism, there is a lower bound on the throat parameter ($\eta$).  The chameleon coupling parameter $\beta$ is constrained from above, with $\beta \lesssim -0.14$ at approximately the $3\sigma$ confidence level for all three data sets. Similar to the previous case, the scalar field parameter $m_0$ is subject to an upper bound, while $m_1 \simeq 0$ for all three data sets. 

For all three data combinations considered here, the model yields $\Delta \chi^{2} < 0$ when compared to the $\Lambda$CDM model. Specifically, for the \setone\ combination, we find $\Delta \chi^{2} = -0.33$, for \settwo\ it is $\Delta \chi^{2} = -0.52$, and for \setthree\ it is $\Delta \chi^{2} = -0.23$. 

An interesting point to note is that the inclusion of the chameleon mechanism does not improve $\Delta \chi^{2}$ for the DBI model; rather, it remains nearly identical to the previous case up to the first decimal place. However, the inclusion of the additional parameter $\beta$ associated with the chameleon mechanism leads to a stronger penalty in $\Delta \mathrm{AIC}$. Consequently, the performance of the model with the chameleon mechanism is slightly worse than that of the model without it when compared to the $\Lambda$CDM scenario.

\begin{figure}
    \centering
    \includegraphics[width=1\linewidth]{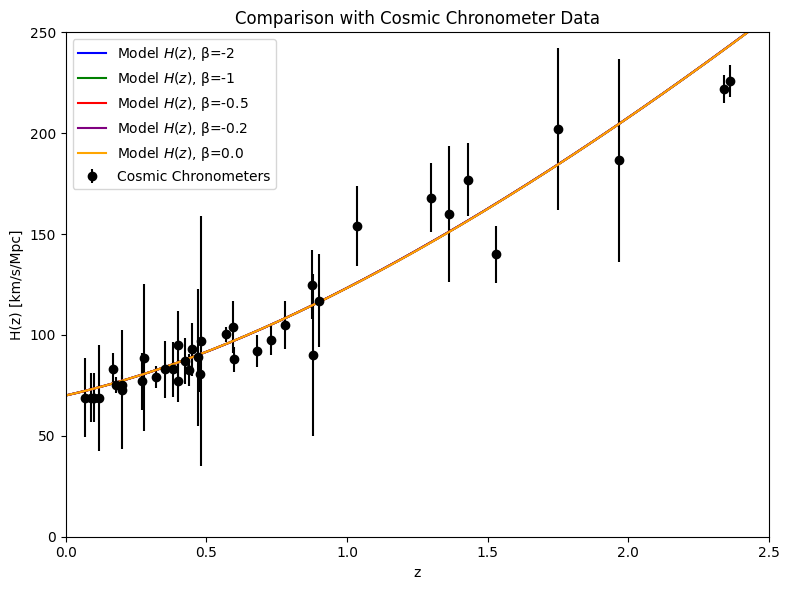}
    \caption{Evolution of 
$H(z)$ versus redshift
$z$ for the DBI model with the chameleon mechanism. Data points from Cosmic Chronometer datasets are shown for comparison. For the choice of the parameters $m_0 =0.1, m_1 = 0.05,\eta=1$ with the current value of the $\phi_0 \simeq 1, \dot{\phi}_0 \simeq 0.1$.}
    \label{fig:H_with}
\end{figure}

\begin{figure}
    \centering
    \includegraphics[width=1\linewidth]{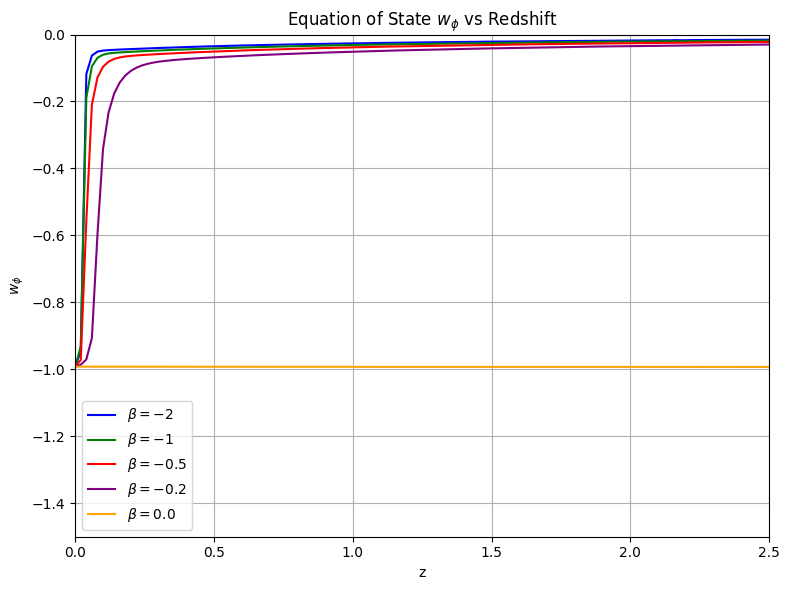}
    \caption{Evolution of the EOS of the dark energy for the DBI model with the chameleon mechanism for different choices of the chameleon coupling parameter $\beta$. For the choice of the parameters $m_0 =0.1, m_1 = 0.05,\eta=1$ with the current value of the $\phi_0 \simeq 1, \dot{\phi}_0 \simeq 0.1$.}
    \label{fig:eos_with}
\end{figure}

In order to evaluate cosmological parameters, we have illustrated the evolution of $H(z)$ as a function of $z$ in Fig. \ref{fig:H_with}. Additionally, observational data from the cosmic chronometer dataset is plotted for comparative purposes. The parameter values chosen for this visualization fall within the range of the posterior distributions derived from the MCMC analysis. Additionally, Fig. \ref{fig:eos_with} illustrates the evolution of the scalar field's equation of state (EoS) utilizing the same set of parameters. The dark energy EoS exhibits quite a different behavior in this scenario compared to the situation where the chameleon mechanism is absent. With the chameleon mechanism in the distant past, the DBI field EoS was zero, mirroring the properties of matter; however, more recently, it has evolved to $w_\phi \approx -1$. This suggests that the DBI field with the chameleon mechanism could potentially offer a unified framework for describing both dark matter and dark energy, which we will explore in more detail in future studies.

\section{conclusion}

The results from recent cosmological observations, such as DESI DR2 and DES Y5, suggest that dark energy might not be constant but rather evolving. Various proposals have been put forward for constructing dynamical dark energy models. In this work, we consider the Dirac-Born-Infeld (DBI) dark energy model and investigate its viability in light of current cosmological observations. We study this model both with and without the inclusion of the chameleon mechanism.

Current cosmological observations, including Pantheon Plus, DES Y5, DESI DR2, and the compressed Planck likelihood, have been used to constrain these models. In this study, we consider an AdS throat with $f(\phi) = \lambda / \phi^4$ and a potential of the form $V(\phi) = m_0^2 \phi^2 + m_1^2 \phi^4$.

A key finding of our analysis is that, for both cases—DBI with and without the chameleon mechanism—the mean value $m_1 \simeq 0$ suggests that the DBI field may lack significant self-interaction. In addition, we find an upper bound on the potential parameter $m_1$. We also observe that the warp parameter satisfies $\eta \geq 0$, while the chameleon coupling parameter is constrained to $\beta \leq 0$. For different choices of the coupling parameter $\beta$, the evolution of the background cosmological parameters remains essentially unchanged; however, the equation of state parameter $w_{\mathrm{DE}}$ tends to lie more within the quintessence region at earlier times. Under the current assumptions for the throat and the potential, we do not observe any crossing of the phantom divide.

We have also performed a statistical model comparison by computing the $\Delta \mathrm{AIC}$ relative to the $\Lambda$CDM model. Although these models provide a slightly better fit to the data than $\Lambda$CDM, the $\Lambda$CDM model remains statistically favored. The chameleon mechanism does not improve $\Delta \chi^{2}$, while the additional parameter $\beta$ increases the $\Delta \mathrm{AIC}$ penalty, rendering the model slightly less favored than its non-chameleon counterpart relative to $\Lambda$CDM.

\section*{Appendix: derivation of the field equation} \label{ap_fieldeq}
From the action (\ref{eq_action}), we derived 
\bea
\Gamma'\, =\,  -\f{1}{2\Gamma} \l( \f{T'}{T}\r) \l( \Gamma^2  - 1\r); 
\eea

\bea
\f{\partial \Gamma}{\partial (\partial_{\nu} \phi)} 
\,=\, \f{g^{\mu \nu}}{\Gamma} \f{\partial_{\mu}\phi}{T}.
\eea
Varying the action with respect to the scalar field. The first term of the Euler-Lagrange equation is 
\be
\f{\pa}{\pa \phi}\l({\mathcal{L}}_{\phi + \m}   \r)  \:=\:  \sqrt{-g} \l[ - \f{T'}{2\Gamma} (\Gamma -1 )^2 \,-\,V'     \r]  \:+ \:   \mathcal{L}_{\m}'    \no
\ee
Then we derive 
\be
\f{\pa {\mathcal{L}}_{\phi+\m}}{\pa(\pa_{\nu} \phi)}  \:=\: - \sqrt{-g} \f{1}{\Gamma} \pa^{\nu} \phi
\ee

\be
\nabla_{\nu}\l[ \f{\pa {\mathcal{L}}_{\phi+\m}}{\pa(\pa_{\nu} \phi)}\r]  \:=\: - \f{\sqrt{-g}}{\Gamma}   \l[ \Box^2 \phi  -  \f{1}{\Gamma}g^{\rho \nu} (\pa_{\rho} \phi) (\pa_{\nu} \Gamma) \r]   \no
\ee
where ${\mathcal{L}}_{\phi +\m}  \equiv    {\mathcal{L}}_{\phi}  + {\mathcal{L}}_{\m} $, 
$\:\;{\pa {\mathcal{L}}_{\m}}/{\pa(\pa_{\nu} \phi)} = 0$,
 $\:\;\nabla_{\nu} \phi  = \partial_{\nu} \phi,\;$ $\;\:\nabla_{\nu} \Gamma^{-1} = - \l(\partial_{\nu} \Gamma \r) / \Gamma^2 \;$  and  $\;\: \Box^2 \phi  \equiv g^{\mu \nu} \nabla_{\mu}\nabla_{\nu} \phi$.

The DBI equation of motion is hence
\be
 \f{\Box^2 \phi }{\Gamma}  \,  -\f{T'}{2\Gamma} \, (\Gamma -1)^2  \,  -  \, \f{1}{\Gamma^2} \, g^{\mu\nu} (\pa_{\mu} \phi )(\pa_{\nu} \Gamma)  \:=\:   \, V'  - \f{\mathcal{L}'_m}{\sqrt{-g}}\,.
\ee

\section*{Acknowledgments} 
 This research project is funded by Mahidol University (via Fundamental Fund: fiscal
 year 2025 by National Science Research and Innovation Fund (NSRF), Thailand). The authors are grateful to Philippe Brax, Carsten van de Bruck, Anne-Christine Davis, Ruth Gregory and David Tong for long-back early discussions related to ideas of this project.

\bibliography{dbichem}

\end{document}